# A Gridless Compressive Sensing Based Channel Estimation for Millimeter Wave MIMO OFDM Systems with One-Bit Quantization

Mahdi Eskandari
Department of Electrical Engineering
Shahed University
Tehran, Iran
m-eskandari@shahed.ac.ir

Hamidreza Bakhshi
Department of Electrical Engineering
Shahed University
Tehran, Iran
bakhshi@shahed.ac.ir

*Abstract*—This paper considers the problem of estimating the sparse mmWave massive multiple input - multiple output (MIMO) OFDM channel from one-bit quantized measurements. Unlike previous grid-based one-bit compressive sensing approaches, we present a gridless convex method to recover sparse channel form one-bit measurements via Binary Atomic Norm Minimization (BiANM) and Reweighted Binary Atomic Norm Minimization (ReBiANM). Simulation results verify the accuracy of the binary and reweighted binary atomic norm minimization techniques.

## I. INTRODUCTION

Recently, the scalating demand for high data transmission, which is one of the key requirements of the fifth generation (5G) wireless communication systems, have trigged and etracted tremendous interests from both academia and industry, for example [1]–[4]. Because of high operating bandwidth in mmWave systems, the sampling rate of the analog to digital converters (ADCs) scales up [5]. For example, at rates above 100 Msamples per second, ADC power consumption increases quadratically with sampling frequency [5]. High precision ADCs with bandwidths sufficient for mmWave systems are either unavailable or may be too costly and power-hungry for portable devices [6]. A one-bit Analog-to-Digital Converter (ADC) is the simplest device for quantization of an analog signal into a digital. It is the least power consuming quantizer, since the power consumption of ADCs grows exponentially with the number of bits needed to represent all quantization levels [7]. Due to the nonlinear nature of quantization, channel estimation with few-bit ADCs is challenging. MIMO channel estimation is more challenging because the linear combination of transmitted signals from different antennas is quantized. There are some prior works focused on mmWave MIMO channel estimation with few-bit ADCs. In [8] and [9], a Least Square (LS) approach used to estimate the MIMO channel. In those works, channel estimation suffers from large estimation error. This is because of that the quantization error treated as additive white Gaussian noise. The works [10]–[12] based their works on Expectation Maximization (EM). EM needs matrix inversion and so many iterations to converge, so those works have high complexity. Approximate Message Passing (AMP) method used for channel estimation in some recent works such [13]–[15]. These works assumed that the channel coefficients follow an IID Gaussian distribution and did not exploit the sparsity nature of mmWave channels. The work [16] propose an AMP-based channel estimation schemes for broadband sparse mmWave MIMO channels with few-bit ADCs. In [17], an adaptive algorithm was proposed to estimate the mmWave channel with one-bit ADCs.

In this paper we consider the problem of recovering a mmWave sparse channel from one-bit quantized measurements. Motivated by the studies on atomic norm, we develop a new gridless one-bit compressive sensing method for channel estimation. We proposed a gridless channel estimator, denoted as Binary Atomic Norm Minimization (BiANM). For further enhancement, we developed a new formulation of BiANM denoted by Reweighted Binary Atomic Norm Minimization (ReBiANM) which is the tradeoff between atomic $\ell_0$ norm minimization which is not convex and NP-hard to solve and atomic $\ell_1$ norm minimization that is convex.

*notations:* $a$ is a scalar, $\mathbf{a}$ is a vector, $\mathbf{A}$ is a matrix, and $\mathcal{A}$ represents a set. For a vector or matrix, the transpose and Hermitian are denoted by $(.)^T$ and $(.)^H$, respectively. $\mathbf{I}_N$ is the identity matrix of size $N \times N$. $\mathbf{A}^{-1}$ and $\text{trace}(\mathbf{A})$ are respectively the inverse and the trace of $\mathbf{A}$. The operation

vec(.) converts a matrix into a vector. $\mathbf{A} \otimes \mathbf{B}$ is the Kronecker product of $\mathbf{A}$ and $\mathbf{B}$. $\mathbf{A} \succeq 0$ means that $\mathbf{A}$ is positive semidefinite (PSD). sign(.) is the signum function applied component wise to the real and imaginary part of the input argument. So, the output of the signum function is one of the elements of the set $\{\pm 1 \pm j\}$. inf{.} is the infimum of the input set. Real and imaginary part of a complex number is denoted by $\Re\{.\}$ and $\Im\{.\}$, respectively. A circularly symmetric complex Gaussian random vector with mean $\boldsymbol{\mu}$ and covariance $\mathbf{C}$ is denoted by $\mathcal{CN}(\boldsymbol{\mu}, \mathbf{C})$. $\mathbb{E}[.]$ denotes expectation. Finally, $\mathbf{a}_K(\beta) = [e^{-j2\pi\beta K'}]^T$ is the complex exponential sequence of length $K$ with $K' = 0, 1, ..., K-1$ and $\beta \in [0,1]$.

## II. SYSTEM AND CHANNEL DESCRIPTION

We consider the uplink of a massive MIMO system. A single-antenna User Equipment (UE) is communicating with a Base Station (BS) with a ULA of $M (M \gg 1)$ antennas. Transmission is performed via OFDM modulation with $N$ subcarriers. The $M \times N$ complex baseband channel with the $(m,n)$-th element correspond to the channel gain of antenna $m$ and subcarrier $n$ is as follows [18]

$$\mathbf{H} = \sum_{l=0}^{L-1} \alpha_l \mathbf{a}_M(\theta_l) \mathbf{a}_N^H(\tau_l) = \sum_{l=0}^{L-1} \alpha_l \mathbf{A}(\Omega_l), \quad (1)$$

where $L \ll \min(M,N)$ is the number of propagation paths, $\alpha_l$ is the gain of the $l$-th path. $\theta_l \in [0,1]$ and $\tau_l \in [0,1]$ is the angle of arrival (AoA) and delay of the $l$-th path, respectively. Also $\mathbf{A}(\Omega) = \mathbf{a}_M(\theta) \mathbf{a}_N^H(\tau)$. With this assumption the received baseband signal is

$$\mathbf{Y} = \mathbf{H}\mathbf{X} + \mathbf{N}, \quad (2)$$

where $\mathbf{X} = \text{diag}(\mathbf{x})$ and $\mathbf{x} \in \mathbb{C}^{N \times 1}$ is the transmitted symbols. $\mathbf{N}$ is additive white Gaussian noise. With one-bit ADC, the receiver will obtain

$$\mathbf{R} = \text{sign}(\mathbf{Y}) = \text{sign}(\mathbf{H}\mathbf{X} + \mathbf{N}), \quad (3)$$

The target is to estimate the channel based on observation $\mathbf{R}$ and the training signal $\mathbf{X}$. The vectorized received signal is

$$\begin{aligned}\mathbf{r} &= \text{vec}(\mathbf{R}) = \text{sign}(\text{vec}(\mathbf{H}\mathbf{X}) + \text{vec}(\mathbf{N})) \\ &\stackrel{(a)}{=} \text{sign}((\mathbf{X}^T \otimes \mathbf{I})\mathbf{h} + \mathbf{n}) \\ &= \text{sign}(\boldsymbol{\Phi}\mathbf{h} + \mathbf{n}),\end{aligned} \quad (4)$$

Where $\quad \mathbf{h} = \text{vec}(\mathbf{H}) = \sum_{l=0}^{L-1} \alpha_l \underbrace{\mathbf{a}_N^H(\tau_l) \otimes \mathbf{a}_M(\theta_l)}_{\mathbf{a}(\omega_l)}$ and (a) follows from the equality $\text{vec}(\mathbf{ABC}) = (\mathbf{C}^T \otimes \mathbf{A})\text{vec}(\mathbf{B})$. The problem in (4) can be analyzed with one-bit compressed sensing framework

$$\overline{\mathbf{R}} = \text{sign}(\mathbf{A}\mathbf{k}), \quad (5)$$

where $\mathbf{k}$ is the sparse vector to be estimated via the measurement vector $\mathbf{A}$. In the channel estimation problem, the matrix $\mathbf{A}$ is $\boldsymbol{\Phi} = (\mathbf{X}^T \otimes \mathbf{I})$ and $\mathbf{k} = \mathbf{h}$. Note that with this binary measurement, one cannot obtain any information about the magnitude of the channel. So, the best we hope to do is to estimate the normalized version of the channel that placed on the unit hypersphere. i.e., $\|\mathbf{h}\|_2 = 1$. Using the sparsity of the channel, we can use the gridless compressed sensing [19] to estimate the channel with the atomic norm minimization concept.

## III. GRIDLESS CHANNEL ESTIMATION BASED ON BINARY ATOMIC NORM MINIMIZATION

Let $\mathcal{A}$ be the set of atoms which construct a signal and defines as follows.

$$\mathcal{A} = \{\mathbf{a}(\omega) \mid \omega \in [0,1] \times [0,1]\}. \quad (6)$$

The atomic $\ell_0$ norm $\|.\|_{\mathcal{A},0}$ of the channel is defend as follows

$$\|\mathbf{h}\|_{\mathcal{A},0} = \inf\{L \mid \mathbf{h} = \sum_{l=0}^{L-1} \alpha_l \mathbf{a}(\omega_l)\}. \quad (7)$$

The $\ell_0$ norm exploits sparsity to the greatest extent possible, but it is not convex and NP-hard to compute and cannot be globally solved with a practical algorithm. The atomic norm $\|.\|_{\mathcal{A}}$ is defined by its unit ball with the convex hull of $\mathcal{A}$ and is the convex relaxation of the atomic $\ell_0$ norm [20], [21].

$$\begin{aligned}\|\mathbf{h}\|_{\mathcal{A}} &= \inf\{\epsilon > 0 \mid \mathbf{h} \in \epsilon\text{conv}(\mathcal{A})\} \\ &= \inf\{\sum_{l=0}^{L} \alpha_l \mid \mathbf{h} = \sum_{l=0}^{L} \alpha_l \mathbf{a}(\omega_l), \mathbf{a} \in \mathcal{A}\}\end{aligned} \quad (8)$$

With the above definitions, the estimation of the channel can be obtained with BiANM.

$$\begin{aligned}\hat{\mathbf{h}}_{\text{BiANM}} = \arg\min \|\mathbf{h}\|_{\mathcal{A}} \\ \text{s.t.} \quad \Re(\mathbf{R}')\Re(\boldsymbol{\Phi}\mathbf{h}) \geq 0 \\ \Im(\mathbf{R}')\Im(\boldsymbol{\Phi}\mathbf{h}) \geq 0 \\ \|\Re(\boldsymbol{\Phi}\mathbf{h})\|_1 + \|\Im(\boldsymbol{\Phi}\mathbf{h})\|_1 = 1,\end{aligned} \quad (9)$$

where first and second constraint followed from the fact that the product of each quantized measurement with the measurement is always non-negative. The inequality is applied element wise and $\mathbf{R}' = \mathrm{diag}(\mathbf{r})$. The last term is the convex relaxation of $\|\mathbf{h}\|_2 = 1$ that turns the optimization problem to the convex form and also prevents the zero solution. As studied in [22] the atomic norm can be computed by semi definite programming (SDP):

$$\|\mathbf{h}\|_{\mathcal{A}} = \arg\min \frac{1}{2MN}\mathrm{trace}(\mathbf{T}_{2D}(\mathbf{u})) + \frac{\delta}{2}$$
$$\text{s.t.} \quad \begin{bmatrix} \mathbf{T}_{2D}(\mathbf{u}) & \mathbf{h} \\ \mathbf{h}^H & \delta \end{bmatrix} \succeq 0, \tag{10}$$

where $\delta \in \mathbb{R}$, $\mathbf{u} \in \mathbb{C}^{MN \times 1}$ and $\mathbf{T}_{2D}(\mathbf{u})$ defined by its first row $\mathbf{u}$ of length $MN$, denotes a two-level Hermitian Toeplitz matrix constructed from the two-level Vandermonde structure of $\mathbf{A}(\mathbf{\Omega})$ [23] With this assumption (10) can be reformulated as

$$\hat{\mathbf{h}}_{\text{BiANM}} = \arg\min \frac{1}{2MN}\mathrm{trace}(\mathbf{T}_{2D}(\mathbf{u})) + \frac{\delta}{2}$$
$$\text{s.t.} \quad \begin{bmatrix} \mathbf{T}_{2D}(\mathbf{u}) & \mathbf{h} \\ \mathbf{h}^H & \delta \end{bmatrix} \succeq 0$$
$$\Re(\mathbf{R}')\Re(\mathbf{\Phi h}) \geq 0 \tag{11}$$
$$\Im(\mathbf{R}')\Im(\mathbf{\Phi h}) \geq 0$$
$$\|\Re(\mathbf{\Phi h})\|_1 + \|\Im(\mathbf{\Phi h})\|_1 = 1.$$

When (11) is solved, the optimal value of $\hat{\mathbf{h}}_{\text{BiANM}}$ which is an estimate of $\mathbf{h}$ will obtain.

As mentioned earlier the optimization problem in (11), however promotes sparsity to the greatest extent possible, it is nonconvex and NP-hard to solve. But for using the sparsity nature of the channel to the great extent, an algorithm should be developed. Reweighted atomic norm Minimization (ReANM) [24] is the solution. With ReANM, the gap between two norms can be mitigated and the sparsity will be enhanced. The detailed implementation approach of ReANM algorithm are given in [24] and the extension to the binary platform is formulated as a Reweighted Binary Atomic Norm Minimization (ReBiANM) as follows.

$$\hat{\mathbf{h}}_{\text{BiReANM}} = \arg\min \frac{1}{2MN}\mathrm{trace}(\mathbf{\Theta}_j \mathbf{T}_{2D}(\mathbf{u})) + \frac{\delta}{2}$$
$$\text{s.t.} \quad \begin{bmatrix} \mathbf{T}_{2D}(\mathbf{u}) & \mathbf{h} \\ \mathbf{h}^H & \delta \end{bmatrix} \succeq 0 \tag{12}$$
$$\Re(\mathbf{R}')\Re(\mathbf{\Phi h}) \geq 0$$
$$\Im(\mathbf{R}')\Im(\mathbf{\Phi h}) \geq 0$$
$$\|\Re(\mathbf{\Phi h})\|_1 + \|\Im(\mathbf{\Phi h})\|_1 = 1.$$

where $\mathbf{\Theta}_j = (\mathbf{T}_{2D}(\mathbf{u}^{(j-1)}) + \zeta \mathbf{I}_{N_r N_t})^{-1}$, $j = 1, 2, ..., J$ and $\mathbf{u}^{(0)}$ equals $\mathbf{u}$ that obtains from (11) and $J$ is the number of iterations. $\zeta$ is a regularization parameter causes the optimization problem in (12) play $\ell_0$ norm minimization a $\zeta \to 0$ and $\ell_1$ norm minimization problem as $\zeta \to \infty$.

IV. SIMULATION RESULTS

In this section, we evaluate the performance of the proposed downlink mmWave channel estimators with one-bit measurement. A BS with $M = 64$ antenna is assumed that is communicating with a single antenna UE with $N = M$ OFDM subcarriers. To get the results, we use Monte-Carlo simulation averaged over many independent realizations. For the ReBiANM algorithm, we initial the value of $\zeta$ equals to 1 and halve $\zeta$ when beginning a new iteration. The iteration number for ReBiANM is set to be $J = 5$. Finally, we used CVX for solving optimization problems.

Since in the mmWave systems we are dealing with noisy measurements, we consider an oversampling regime in order to compensate for the potential errors in the measurements due to the high noise levels. In this situation 5 samples are taken with the same measurement but different noise realizations. The final measurement is computed after a majority vote decision.

The performance metric is the normalized mean square error (MSE), given by $\mathbb{E}\left[\dfrac{\|\hat{\mathbf{h}} - \mathbf{h}\|^2}{\|\mathbf{h}\|^2}\right]$ where $\hat{\mathbf{h}}$ is the vectorized estimate of vectorized channel $\mathbf{h}$.

Fig. 1, compares grisless algorithms as a function of Signal to Noise Ratio (SNR). Also, the oversampling methods act better than corresponding no oversampling case in low SNR regime. The reason is that in high SNR's the effect of noise is lower than low SNR's. So, the NMSE in high SNR is similar to no oversampling case.

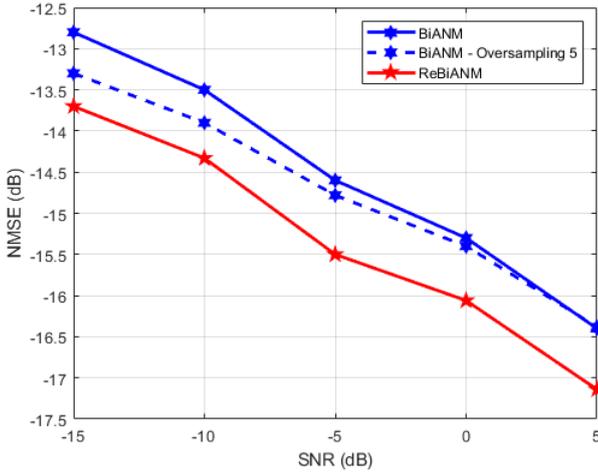

Fig. 1. Performance comparison of different algorithms when there are $L = 3$ paths of the channel.

## V. CONCLUSION

In this paper, we studied the problem of channel estimation in sparse and continuous regime from one-bit measurements with BiANM method. Also, we provide ReBiANM algorithm to solve BiANM with more accuracy. Numerical experiments verified the accuracy of proposed methods.